\documentstyle[10pt,epsf,epsfig,dp_delphititle]{dp_delphi}
\makeindex
\def\DpPaperGroup{EP}
\def\DpPaperRef{99--03}
\def\DpDate{7 January 1999}
\def\DpAuthors{DELPHI Collaboration}
\def\DpSubmit{(Submitted to Physics Letters B)}
\def\DpTitle{{The Scale Dependence of the Hadron Multiplicity 
in Quark and Gluon Jets and a Precise Determination of
\boldmath ${\rm  C_A/C_F}$}}
\begin{document}
\makeatletter
\newcount\@tempcntc
\def\@citex[#1]#2{\if@filesw\immediate\write\@auxout{\string\citation{#2}}\fi
  \@tempcnta\z@\@tempcntb\m@ne\def\@citea{}\@cite{\@for\@citeb:=#2\do
    {\@ifundefined
       {b@\@citeb}{\@citeo\@tempcntb\m@ne\@citea\def\@citea{,}{\bf ?}\@warning
       {Citation `\@citeb' on page \thepage \space undefined}}%
    {\setbox\z@\hbox{\global\@tempcntc0\csname b@\@citeb\endcsname\relax}%
     \ifnum\@tempcntc=\z@ \@citeo\@tempcntb\m@ne
       \@citea\def\@citea{,}\hbox{\csname b@\@citeb\endcsname}%
     \else
      \advance\@tempcntb\@ne
      \ifnum\@tempcntb=\@tempcntc
      \else\advance\@tempcntb\m@ne\@citeo
      \@tempcnta\@tempcntc\@tempcntb\@tempcntc\fi\fi}}\@citeo}{#1}}
\def\@citeo{\ifnum\@tempcnta>\@tempcntb\else\@citea\def\@citea{,}%
  \ifnum\@tempcnta=\@tempcntb\the\@tempcnta\else
   {\advance\@tempcnta\@ne\ifnum\@tempcnta=\@tempcntb \else \def\@citea{--}\fi
    \advance\@tempcnta\m@ne\the\@tempcnta\@citea\the\@tempcntb}\fi\fi}
 
\makeatother
\begin{titlepage}
\pagenumbering{roman}
\CERNpreprint{\DpPaperGroup}{\DpPaperRef} 
\date{{\small\DpDate}} 
\title{\DpTitle} 
\address{\DpAuthors} 
\begin{shortabs} 
\noindent
%
\newcommand{\cacf} {\ifmmode{C_A/C_F}       \else{$C_A/C_F$}\fi}

\noindent

Data collected at the Z resonance using the DELPHI detector at LEP
are used to determine the charged hadron multiplicity in gluon and quark
jets as a function of a transverse momentum-like scale. The colour
factor ratio, \cacf, is directly observed in the increase of
multiplicities with that scale.
 The smaller than expected  multiplicity ratio in gluon to
quark jets is understood by differences in the hadronization of
the leading quark or gluon. From the dependence of the charged hadron
multiplicity on the opening angle in symmetric three-jet events
the colour factor ratio is measured to be: 
$$
\frac{C_A}{C_F} = 2.246 \pm 0.062~(stat.) \pm 0.080~(syst.) \pm 0.095~(theo.) 
$$
\end{shortabs}
\vfill
\begin{center}
\DpSubmit \ 
\end{center}
\vfill
\clearpage
\headsep 10.0pt
\addtolength{\textheight}{10mm}
\addtolength{\footskip}{-5mm}
\begingroup
%
\newcommand{\DpName}[2]{\hbox{#1$^{\ref{#2}}$},\hfill}
\newcommand{\DpNameTwo}[3]{\hbox{#1$^{\ref{#2},\ref{#3}}$},\hfill}
\newcommand{\DpNameThree}[4]{\hbox{#1$^{\ref{#2},\ref{#3},\ref{#4}}$},\hfill}
\newskip\Bigfill \Bigfill = 0pt plus 1000fill
\newcommand{\DpNameLast}[2]{\hbox{#1$^{\ref{#2}}$}\hspace{\Bigfill}}
%
\footnotesize
\noindent
\DpName{P.Abreu}{LIP}
\DpName{W.Adam}{VIENNA}
\DpName{T.Adye}{RAL}
\DpName{P.Adzic}{DEMOKRITOS}
\DpName{I.Ajinenko}{SERPUKHOV}
\DpName{Z.Albrecht}{KARLSRUHE}
\DpName{T.Alderweireld}{AIM}
\DpName{G.D.Alekseev}{JINR}
\DpName{R.Alemany}{VALENCIA}
\DpName{T.Allmendinger}{KARLSRUHE}
% ...
\normalsize
\endgroup
\titlefoot{Department of Physics and Astronomy, Iowa State
     University, Ames IA 50011-3160, USA
    \label{AMES}}
\titlefoot{Physics Department, Univ. Instelling Antwerpen,
     Universiteitsplein 1, BE-2610 Wilrijk, Belgium \\
     \indent~~and IIHE, ULB-VUB,
     Pleinlaan 2, BE-1050 Brussels, Belgium \\
     \indent~~and Facult\'e des Sciences,
     Univ. de l'Etat Mons, Av. Maistriau 19, BE-7000 Mons, Belgium
    \label{AIM}}
\titlefoot{Physics Laboratory, University of Athens, Solonos Str.
     104, GR-10680 Athens, Greece
    \label{ATHENS}}
\titlefoot{Department of Physics, University of Bergen,
     All\'egaten 55, NO-5007 Bergen, Norway
    \label{BERGEN}}
\titlefoot{Dipartimento di Fisica, Universit\`a di Bologna and INFN,
     Via Irnerio 46, IT-40126 Bologna, Italy
    \label{BOLOGNA}}
\titlefoot{Centro Brasileiro de Pesquisas F\'{\i}sicas, rua Xavier Sigaud 150,
     BR-22290 Rio de Janeiro, Brazil \\
     \indent~~and Depto. de F\'{\i}sica, Pont. Univ. Cat\'olica,
     C.P. 38071 BR-22453 Rio de Janeiro, Brazil \\
     \indent~~and Inst. de F\'{\i}sica, Univ. Estadual do Rio de Janeiro,
     rua S\~{a}o Francisco Xavier 524, Rio de Janeiro, Brazil
    \label{BRASIL}}
\titlefoot{Comenius University, Faculty of Mathematics and Physics,
     Mlynska Dolina, SK-84215 Bratislava, Slovakia
    \label{BRATISLAVA}}
\titlefoot{Coll\`ege de France, Lab. de Physique Corpusculaire, IN2P3-CNRS,
     FR-75231 Paris Cedex 05, France
    \label{CDF}}
\titlefoot{CERN, CH-1211 Geneva 23, Switzerland
    \label{CERN}}
\titlefoot{Institut de Recherches Subatomiques, IN2P3 - CNRS/ULP - BP20,
     FR-67037 Strasbourg Cedex, France
    \label{CRN}}
\titlefoot{Institute of Nuclear Physics, N.C.S.R. Demokritos,
     P.O. Box 60228, GR-15310 Athens, Greece
    \label{DEMOKRITOS}}
\titlefoot{FZU, Inst. of Phys. of the C.A.S. High Energy Physics Division,
     Na Slovance 2, CZ-180 40, Praha 8, Czech Republic
    \label{FZU}}
\titlefoot{Dipartimento di Fisica, Universit\`a di Genova and INFN,
     Via Dodecaneso 33, IT-16146 Genova, Italy
    \label{GENOVA}}
\titlefoot{Institut des Sciences Nucl\'eaires, IN2P3-CNRS, Universit\'e
     de Grenoble 1, FR-38026 Grenoble Cedex, France
    \label{GRENOBLE}}
\titlefoot{Helsinki Institute of Physics, HIP,
     P.O. Box 9, FI-00014 Helsinki, Finland
    \label{HELSINKI}}
\titlefoot{Joint Institute for Nuclear Research, Dubna, Head Post
     Office, P.O. Box 79, RU-101 000 Moscow, Russian Federation
    \label{JINR}}
\titlefoot{Institut f\"ur Experimentelle Kernphysik,
     Universit\"at Karlsruhe, Postfach 6980, DE-76128 Karlsruhe,
     Germany
    \label{KARLSRUHE}}
\titlefoot{Institute of Nuclear Physics and University of Mining and Metalurgy,
     Ul. Kawiory 26a, PL-30055 Krakow, Poland
    \label{KRAKOW}}
\titlefoot{Universit\'e de Paris-Sud, Lab. de l'Acc\'el\'erateur
     Lin\'eaire, IN2P3-CNRS, B\^{a}t. 200, FR-91405 Orsay Cedex, France
    \label{LAL}}
\titlefoot{School of Physics and Chemistry, University of Lancaster,
     Lancaster LA1 4YB, UK
    \label{LANCASTER}}
\titlefoot{LIP, IST, FCUL - Av. Elias Garcia, 14-$1^{o}$,
     PT-1000 Lisboa Codex, Portugal
    \label{LIP}}
\titlefoot{Department of Physics, University of Liverpool, P.O.
     Box 147, Liverpool L69 3BX, UK
    \label{LIVERPOOL}}
\titlefoot{LPNHE, IN2P3-CNRS, Univ.~Paris VI et VII, Tour 33 (RdC),
     4 place Jussieu, FR-75252 Paris Cedex 05, France
    \label{LPNHE}}
\titlefoot{Department of Physics, University of Lund,
     S\"olvegatan 14, SE-223 63 Lund, Sweden
    \label{LUND}}
\titlefoot{Universit\'e Claude Bernard de Lyon, IPNL, IN2P3-CNRS,
     FR-69622 Villeurbanne Cedex, France
    \label{LYON}}
\titlefoot{Univ. d'Aix - Marseille II - CPP, IN2P3-CNRS,
     FR-13288 Marseille Cedex 09, France
    \label{MARSEILLE}}
\titlefoot{Dipartimento di Fisica, Universit\`a di Milano and INFN,
     Via Celoria 16, IT-20133 Milan, Italy
    \label{MILANO}}
\titlefoot{Niels Bohr Institute, Blegdamsvej 17,
     DK-2100 Copenhagen {\O}, Denmark
    \label{NBI}}
\titlefoot{NC, Nuclear Centre of MFF, Charles University, Areal MFF,
     V Holesovickach 2, CZ-180 00, Praha 8, Czech Republic
    \label{NC}}
\titlefoot{NIKHEF, Postbus 41882, NL-1009 DB
     Amsterdam, The Netherlands
    \label{NIKHEF}}
\titlefoot{National Technical University, Physics Department,
     Zografou Campus, GR-15773 Athens, Greece
    \label{NTU-ATHENS}}
\titlefoot{Physics Department, University of Oslo, Blindern,
     NO-1000 Oslo 3, Norway
    \label{OSLO}}
\titlefoot{Dpto. Fisica, Univ. Oviedo, Avda. Calvo Sotelo
     s/n, ES-33007 Oviedo, Spain
    \label{OVIEDO}}
\titlefoot{Department of Physics, University of Oxford,
     Keble Road, Oxford OX1 3RH, UK
    \label{OXFORD}}
\titlefoot{Dipartimento di Fisica, Universit\`a di Padova and
     INFN, Via Marzolo 8, IT-35131 Padua, Italy
    \label{PADOVA}}
\titlefoot{Rutherford Appleton Laboratory, Chilton, Didcot
     OX11 OQX, UK
    \label{RAL}}
\titlefoot{Dipartimento di Fisica, Universit\`a di Roma II and
     INFN, Tor Vergata, IT-00173 Rome, Italy
    \label{ROMA2}}
\titlefoot{Dipartimento di Fisica, Universit\`a di Roma III and
     INFN, Via della Vasca Navale 84, IT-00146 Rome, Italy
    \label{ROMA3}}
\titlefoot{DAPNIA/Service de Physique des Particules,
     CEA-Saclay, FR-91191 Gif-sur-Yvette Cedex, France
    \label{SACLAY}}
\titlefoot{Instituto de Fisica de Cantabria (CSIC-UC), Avda.
     los Castros s/n, ES-39006 Santander, Spain
    \label{SANTANDER}}
\titlefoot{Dipartimento di Fisica, Universit\`a degli Studi di Roma
     La Sapienza, Piazzale Aldo Moro 2, IT-00185 Rome, Italy
    \label{SAPIENZA}}
\titlefoot{Inst. for High Energy Physics, Serpukov
     P.O. Box 35, Protvino, (Moscow Region), Russian Federation
    \label{SERPUKHOV}}
\titlefoot{J. Stefan Institute, Jamova 39, SI-1000 Ljubljana, Slovenia
     and Laboratory for Astroparticle Physics,\\
     \indent~~Nova Gorica Polytechnic, Kostanjeviska 16a, SI-5000 Nova Gorica, Slovenia, \\
     \indent~~and Department of Physics, University of Ljubljana,
     SI-1000 Ljubljana, Slovenia
    \label{SLOVENIJA}}
\titlefoot{Fysikum, Stockholm University,
     Box 6730, SE-113 85 Stockholm, Sweden
    \label{STOCKHOLM}}
\titlefoot{Dipartimento di Fisica Sperimentale, Universit\`a di
     Torino and INFN, Via P. Giuria 1, IT-10125 Turin, Italy
    \label{TORINO}}
\titlefoot{Dipartimento di Fisica, Universit\`a di Trieste and
     INFN, Via A. Valerio 2, IT-34127 Trieste, Italy \\
     \indent~~and Istituto di Fisica, Universit\`a di Udine,
     IT-33100 Udine, Italy
    \label{TU}}
\titlefoot{Univ. Federal do Rio de Janeiro, C.P. 68528
     Cidade Univ., Ilha do Fund\~ao
     BR-21945-970 Rio de Janeiro, Brazil
    \label{UFRJ}}
\titlefoot{Department of Radiation Sciences, University of
     Uppsala, P.O. Box 535, SE-751 21 Uppsala, Sweden
    \label{UPPSALA}}
\titlefoot{IFIC, Valencia-CSIC, and D.F.A.M.N., U. de Valencia,
     Avda. Dr. Moliner 50, ES-46100 Burjassot (Valencia), Spain
    \label{VALENCIA}}
\titlefoot{Institut f\"ur Hochenergiephysik, \"Osterr. Akad.
     d. Wissensch., Nikolsdorfergasse 18, AT-1050 Vienna, Austria
    \label{VIENNA}}
\titlefoot{Inst. Nuclear Studies and University of Warsaw, Ul.
     Hoza 69, PL-00681 Warsaw, Poland
    \label{WARSZAWA}}
\titlefoot{Fachbereich Physik, University of Wuppertal, Postfach
     100 127, DE-42097 Wuppertal, Germany
    \label{WUPPERTAL}}
\titlefoot{On leave of absence from IHEP Serpukhov
    \label{MILAN-SERPOU}}
\titlefoot{Now at University of Florida
    \label{FLORIDA}}
\addtolength{\textheight}{-10mm}
\addtolength{\footskip}{5mm}
\clearpage
\headsep 30.0pt
\end{titlepage}
%
\pagestyle{heading} 
\pagenumbering{arabic} 
\setcounter{footnote}{0} %
\large
%
\newcommand{\cacf} {\ifmmode{C_A/C_F}       \else{$C_A/C_F$}\fi}
\newcommand{\lam}  {\ifmmode{\Lambda_{QCD}} \else{$\Lambda_{QCD}$}\fi}
\newcommand{\epem} {\ifmmode{e^+e^-}        \else{$e^+e^-$}\fi}
\newcommand{\qqbar} {\ifmmode{\rm q\bar{q}}        \else{$\rm q\bar{q}$}\fi}

\section{Introduction}
The gauge symmetry underlying the Lagrangian of an
interaction directly determines the
relative coupling of the vertices of the participating elementary fields.
A comparison of the properties of quark and gluon jets, which are linked
to the quark and gluon couplings, therefore implies a direct and intuitive
test of Quantum Chromodynamics, QCD, the gauge theory of the strong interaction.

Hadron production can be described via a so-called parton
shower, a chain of successive bremsstrahlung processes, followed by
hadron formation which cannot be described perturbatively. 
As bremsstrahlung is directly
proportional to the coupling of the radiated vector boson to the
radiator, the ratio of the radiated gluon multiplicity from a
gluon and quark source is  expected to be asymptotically equal to
the ratio of the QCD colour factors: $C_A/C_F=9/4$~\cite{brodskygunion}. 
As the radiated gluons give rise to the production of hadrons, 
the increased radiation from
gluons should be reflected in a higher hadron multiplicity
and also in a stronger scaling violation of
the gluon fragmentation function
\cite{splitting_paper,scaling_vancouver}.

It was however noted already in the first paper
comparing the multiplicities from gluons and quarks \cite{brodskygunion}
that this prediction does not immediately apply to the 
observed charged hadron multiplicities at finite energy
as this  is also influenced by differences of the fragmentation
of the primary quark or gluon.
These differences must be present 
because quarks
are valence particles of the hadrons whereas gluons are not.
This is most clearly evident from the behaviour of the gluon
fragmentation function to charged hadrons at large scaled  momentum
where it is suppressed by about one order of magnitude compared to the
quark fragmentation function \cite{scaling_vancouver}.
This suppression also causes a
higher multiplicity to be expected from very low energy quark
jets
compared to gluon jets.
Moreover, as low momentum, large wavelength gluons cannot resolve a hard
radiated gluon from the initial quark-antiquark pair in the early phase of an
event, soft radiation
and correspondingly the production of low energy hadrons
is further suppressed \cite{stringeffect,ochs,oliver}
compared to the naive expectation.
In a previous publication \cite{splitting_paper} it has
been shown that a reduction of the
primary splittings 
of gluons compared to the perturbative
expectation is indeed
responsible for the observed small hadron multiplicity ratio
between gluon and quark jets.

If heavy quark jets are also included in the comparison, a further
reduction of the multiplicity ratio is evident due to the high number
of particles from the decays of the primary heavy particles.

Furthermore, the definition of quark and gluon jets in three-jet events in
\epem\ annihilation uses jet
algorithms which combine hadrons to make jets.
Low energy particles at large angles with respect to
the original parton direction are
likely to be assigned to a different jet.
As gluon jets are initially wider than quark jets this presumably
leads to a loss of multiplicity for gluon jets and a corresponding
gain for quark jets.

The effects discussed lead to a ratio between the 
charged hadron multiplicities 
from gluon and quark jets being smaller than 
the ratio between gluon radiation from gluons and from quarks.
So far these effects have mainly been neglected in experimental and
more elaborate theoretical investigations \cite{theorie_qg}.
However, as we will show in this paper,
at current energies these non-perturbative effects are still important
and need to be considered in a proper test of the prediction
\cite{brodskygunion} that the radiated gluon-to-quark multiplicity ratio
is equal to the colour factor ratio.

The stronger radiation from gluons is expected to become
directly evident from
a stronger increase of the gluon jet multiplicity with the relevant 
energy scale as compared to quark jets.
In this way the size of the non-perturbative terms can also be directly
estimated from the quark and gluon jet multiplicity at very small scales.
A scale dependence of quark and gluon properties was first
demonstrated in \cite{gluon_paper_1} with the jet energy as the intuitive
scale.
This result was later confirmed by other measurements
\cite{qg_edep_measurments,cleo,opal}
and has recently been extended to a transverse momentum-like scale
\cite{aleph_qgmult}.

A study of the {\it total} charged multiplicity of symmetric three-jet events as function
of the internal scales of the event avoids some of the complications mentioned
above.
A novel precision measurement
of the colour factor ratio \cacf\ can be performed by combining these data
with a Modified Leading Log Approximation 
(MLLA) prediction of the three-jet event multiplicity 
\cite{DKT}
which includes coherence of soft gluon radiation.

This letter is based on a data analysis which is similar to that
presented in previous papers \cite{splitting_paper,gluon_paper_1}.
We therefore have restricted the experimental discussion in
section \ref{experim} to the relevant differences  with respect to
these papers.
More detailed information can also be found in \cite{salva,martin}.
In section \ref{sect:cmp_result} 
the ratio of the slopes of the mean hadron multiplicities 
in gluon and quark jets with scale is shown to
be determined by the colour factor ratio \cacf.
In order to describe the data with the perturbative QCD expectations
it is necessary to introduce additional non-perturbative offsets. 
This analysis is intended to be mainly qualitative and in many
aspects it is similar to previous analyses.
Then 
in section \ref{sect:3jet_result}
a precision measurement of the colour factor ratio from symmetric
three-jet events is discussed and an estimate for the difference of
non-perturbative contributions to the quark and gluon jet multiplicity is given.
Finally we summarize and conclude.

\section{Data Analysis  \label{experim}}
\begin{figure}[htbp]
\begin{center}
\epsfig{file=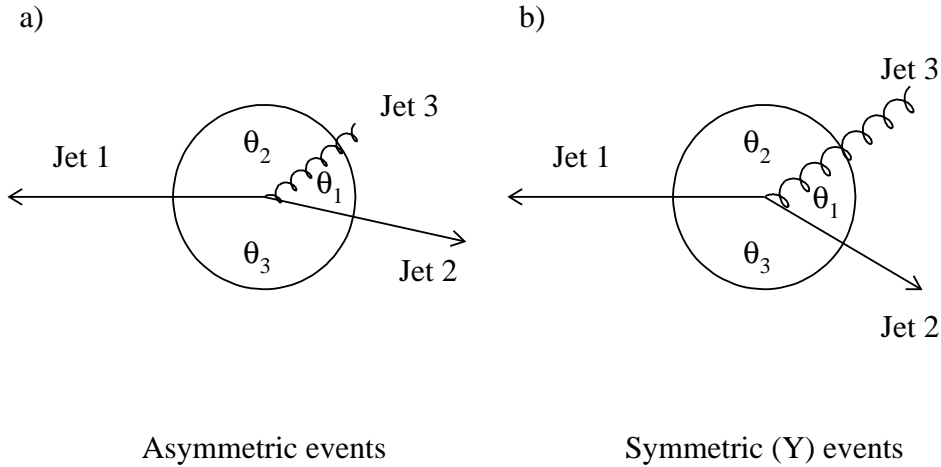,width=14.cm}
\caption[]
{
\label{eventtopol}
Definition of event topologies and angles used throughout this
analysis. The length of the jet lines indicates the energies.
In the symmetric (Y) events (see Fig. \ref{eventtopol}b)) 
$\theta_2\approx\theta_3$.
}
\end{center}
\end{figure}

The analysis presented in this letter uses the full hadronic
data set collected with the DELPHI detector (described in
\cite{delphi_nim_performance}) 
at Z energies in the years 1992 to 1995.
The cuts applied to charged and neutral particles and to events in order
to select hadronic Z decays are identical to those given in
\cite{splitting_paper} for the $q\bar{q}g$ analysis and to
\cite{gluon_paper_1} for the $q\bar{q}\gamma$ analysis.
For the comparison of gluon and quark jets,
three-jet events are clustered using the Durham algorithm
\mbox{\cite{durham_algo}.}
In addition it was required that the angles, $\theta_{2,3}$,
between the low-energy jets
and the leading jet are in the range from $100^{\circ}$ to $170^{\circ}$
(see Fig. \ref{eventtopol}a)).
Within this sample, events are called symmetric if $\theta_{2}$ 
and $\theta_{3}$ are
equal within some analysis-dependent tolerance.
The leading jet is not used in the gluon or quark jet analysis.

The identification of gluon jets by anti-tagging of heavy quark jets
is identical to that described in \cite{splitting_paper,gluon_paper_1}.
Quark jets are taken from $q\bar{q}g$ events which have been depleted in  
b-quark events using an impact parameter technique.
In order to achieve multiplicities of pure quark and gluon jet samples, 
the data have
been corrected using purities from simulated events generated with
JETSET~7.3~\cite{pythia} with parameters set as given in \cite{tuningpaper}.
This is justified by the good agreement between data and simulation.
Furthermore the model independent techniques described in
\cite{splitting_paper} for symmetric events 
(see Fig. \ref{eventtopol}b))
give results largely compatible
with those obtained with the simulation correction \cite{martin}.
The effects of the finite resolution and acceptance of the
detector 
and of the cuts applied 
are corrected for by using a full simulation of the DELPHI detector
\cite{delphi_nim_performance}.

The correction for the remaining b-quark events in the $q\bar{q}g$ sample
does not influence the
slope of the measured multiplicity with scale, but only leads to a
shift of its absolute value.

In the simulation, quark and gluon jets are identified at ``parton
level''.
The partons entering the fragmentation of a three-jet event are clustered 
into three jets using the Durham algorithm.
Then for each parton jet, the number of quarks and antiquarks
are summed where primary quarks contribute with
weight $+1$ and antiquarks with the weight $-1$.
Other quarks and gluons are assigned the weight 0.
These sums are expected to yield $+1$ for quarks jets, $-1$ for anti-quark
jets and 0 for gluon jets.
The small amount of events not showing this expected pattern
of \mbox{$(+1,-1,0)$} was discarded.
Finally, the parton jets were mapped to the jets at the hadron level by
requiring the sum of angles between the parton and hadron jets to be
minimal. 
Events exceeding a maximum angle between the parton and jet directions
were also rejected. 
At large opening angles the influence of these rejections is found
to be about 3\% increasing at low opening angles.

The gluon jet purities vary from 95\% for low energy
gluons to 46\% for the highest energy gluons.
The few bins with lower purities have been excluded from the analysis.
The quark purities range from 43\% to 81\%.

For the analysis of the multiplicity of
symmetric three-jet events, all events were forced
to three jets using the Durham algorithm without a minimal $y_{cut}$.
The angles between the jets were
then used to rescale the jet momenta to the centre-of-mass
energy as described in \cite{gluon_paper_1}. Symmetric events were
selected by demanding
that $\theta_2$ be equal to $\theta_3$ within $2\epsilon$.
Here $\epsilon$ is half the angular bin width of $\theta_1$ taken 
to be $3^\circ$.
The analysis has been performed for events of all flavours as well as for
b-depleted events. In both cases the measured multiplicity was corrected
for track losses due to detector effects and cuts applied. The correction
factor was  calculated as ratio
of generated over accepted multiplicity using simulated events.
It varies smoothly, from 1.25 at small $\theta_1$ to 1.32 at large $\theta_{1}$.

\section{Results \label{results}}
\subsection{Comparison of Multiplicities in Gluon and Quark Jets 
\label{sect:cmp_result}}
In order to determine a scale dependence, the scale underlying the
physics process needs to be specified.
The actual physical scale is necessarily proportional to any variation
of an outer scale like the centre-of-mass energy.
As usually only the relative change in scale matters, this outer scale
can therefore be used instead of the physical scale.
For this analysis the situation is different.
The jets entering the analysis stem from Z decays and thus from a fixed
centre-of-mass energy.
So the relevant scales have to be determined from the properties of
the jets and the event topology.
From the above discussion the scale has to be proportional
to the jet energy because this quantity scales with the
energy in the centre-of-mass system for similar events.
Studies of hadron production in processes with non-trivial topology
have shown that the characteristics of the parton cascade prove to
depend mainly on the hardness of the process producing the jet
\cite{stringeffect,pertQCD}:
\begin{equation}
\kappa = E_{jet} \sin \frac{\theta}{2}~~~.
\label{eqn:kappadef}
\end{equation}
$E_{jet}$ is the energy of the jet and $\theta$ its angle to
the closest jet.
This scale definition corresponds to the beam energy in two-jet events.
It is similar to the transverse momentum of the jet and also
related to $\sqrt{y_{cut}}$ as used by the jet algorithms.
It is also used as the scale in the calculation of the energy dependence of the
hadron multiplicity in $e^+e^-$ annihilation
\cite{webber,colliderphysics}
to take into account the leading effect of coherence.
It should, however, be noted that several scales
may be relevant in multi-jet events. 
Hence using $\kappa$ is an approximation.
A similar scale, namely the geometric mean of the scales of the gluon
jet with respect to both quark jets while 
using Eqn.~\ref{eqn:kappadef} for the quark jets, 
has recently been used in a study of
quark and gluon jet multiplicities \cite{aleph_qgmult}.

As stated in the introduction we want to gain information on the
relative colour charges of quarks and gluons from the 
rate of change of the
multiplicities with scale.
Assuming the validity of the perturbative QCD prediction, the ratio of the
charged multiplicities of gluon and quark jets, 
$N_{gluon}/N_{quark}$, has to approach a constant value
(approximately the colour factor ratio)  at large scale.
This trivially implies that the ratio of the slopes of quark
and gluon jet multiplicities also approaches the same limit.
This fact is a direct consequence of de l'H{\^o}pital's rule
\cite{lhopital} and is also
directly evident from the linearity of the derivative:
\begin{equation}
\mbox{at large scale:} ~~~ N_{gluon}(\kappa)=C \cdot N_{quark }(\kappa) ~~~\rightarrow~~~
\frac{dN_{gluon}/d\kappa}{dN_{quark}/d\kappa} = C~~~,
\label{eqn:trivialitaet}
\end{equation}
i.e. the QCD prediction for the ratio of multiplicities
applies equally well to the ratio of the slopes of the multiplicities.
In fact it is to be expected that the slope ratio is closer to the QCD
prediction than the multiplicity ratio as it should be less affected by
non-perturbative effects.

This effect has been cross-checked using the HERWIG model~\cite{herwig}
which allows the number of colours to be changed 
and thus by SU(n) group relations,
the colour factor ratio \mbox{$C_A/C_F$.}
The predictions of HERWIG are found to follow  directly
the expectation of the right hand side of Eqn.~\ref{eqn:trivialitaet}.
This has also been confirmed in a recent theoretical calculation of this
quantity \cite{eden} in the framework of the dipole model.

\begin{figure}[t]
\begin{center}

\begin{minipage}[t]{224pt}
\epsfig{file=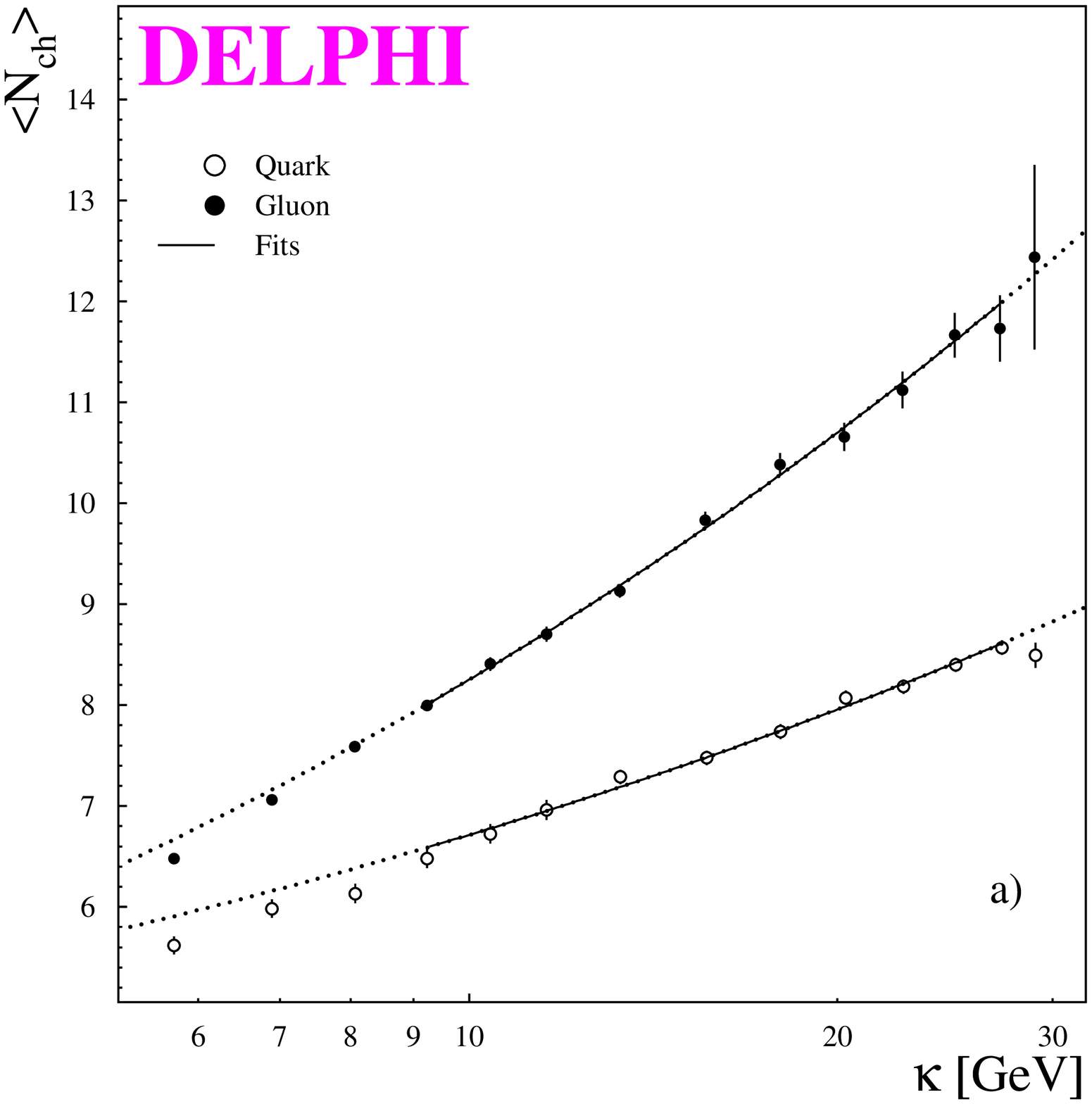,width=220pt}
\end{minipage}
\hfill
\begin{minipage}[t]{224pt}
\epsfig{file=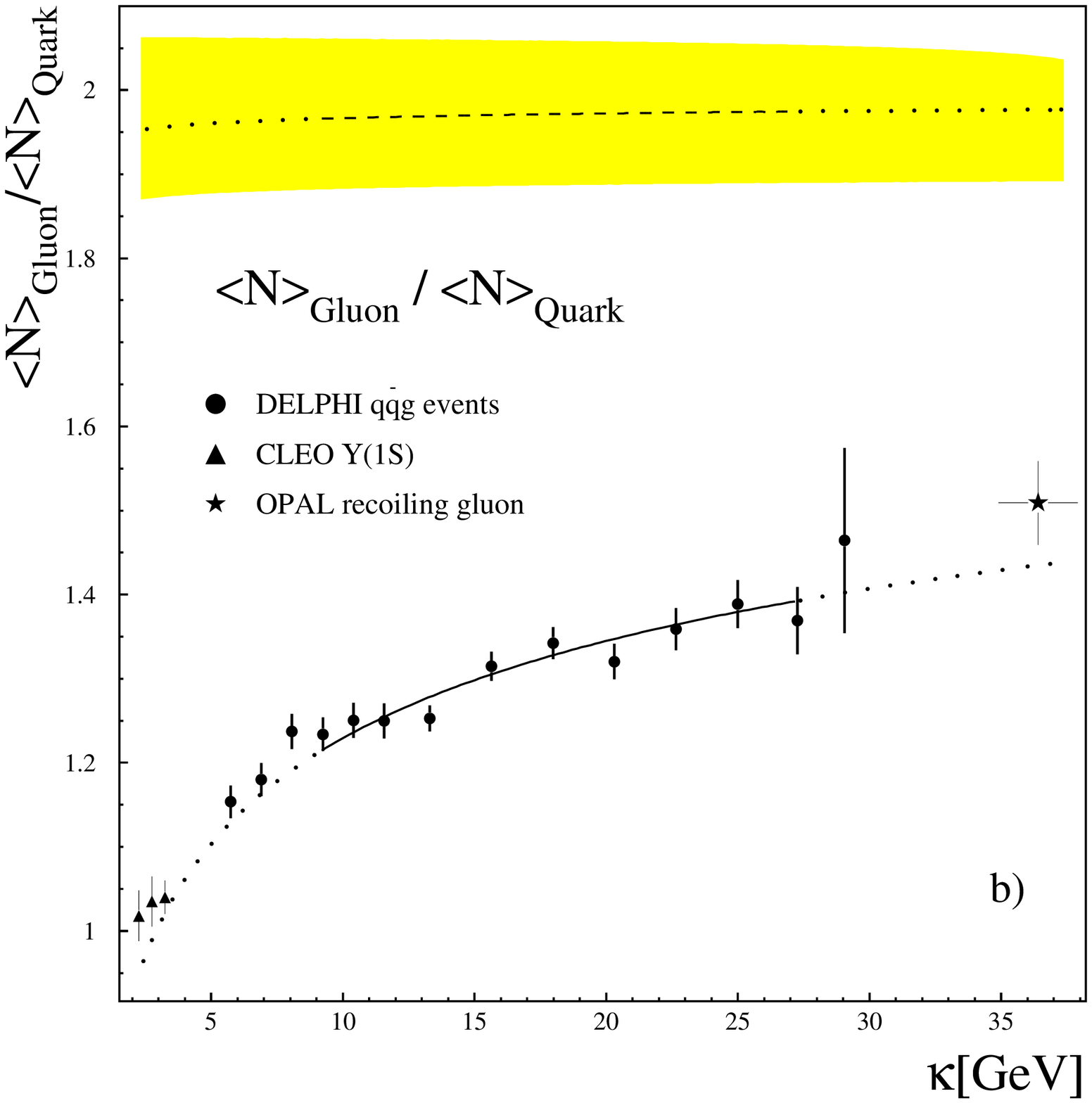,width=220pt}
\end{minipage}

\caption[]
{
\label{qgcmp1}
a) Average charged particle multiplicity for light quark and gluon jets
as function of $\kappa$ fitted with Eqn.~\ref{eqn:offset_ansatz};
b) ratio of the gluon to quark jet multiplicity;
the full line shows the ratio of the functions fitted to the data in 
a), the dashed curve is the ratio of the slopes of the fits in a). 
All curves are extrapolated to the edges of the plot by the dotted
lines.
Also included are measurements of the multiplicity ratio of some other
experiments \cite{cleo,opal}.
The grey band shown with the slope ratio indicates the error
estimated by varying all fit parameters within their errors.
}
\end{center}
\end{figure}

Fig. \mbox{\ref{qgcmp1}a)} shows the multiplicity in quark and gluon
jets as a function of the hardness scale $\kappa$.
For both multiplicities an approximately logarithmic increase with
$\kappa$ is observed which is
about twice as big for gluon jets as for quark jets,
thus already strikingly confirming the QCD prediction.

A stronger increase of the gluon jet multiplicity 
was already noted in a previous paper
\cite{gluon_paper_1}, where the jet energy was chosen as scale.
Meanwhile this observation has been confirmed also by other
measurements \cite{qg_edep_measurments,cleo,opal} and has been extended
to different scales \cite{aleph_qgmult}.
Fragmentation models (not shown) predict an increase of the
multiplicities which is in good agreement with the data.

In order to obtain quantitative information from the data shown in 
Fig.~\mbox{\ref{qgcmp1}a)},
the following ansatz was fitted to the data:
\pagebreak
\begin{eqnarray}
<N_q>(\kappa) &=& N_0^q + N_{pert}(\kappa)   \nonumber \\
<N_g>(\kappa) &=& N_0^g + N_{pert}(\kappa) \cdot r(\kappa)
\label{eqn:offset_ansatz}
\end{eqnarray}
Here $N_0^{q,g}$ are 
non-perturbative terms introduced to account for
the differences in the fragmentation of the leading quark or gluon as discussed
in detail in the introduction.
These terms are assumed to be constant.
$N_{pert}$ is the perturbative prediction for the hadron multiplicity
as given in \cite{webber}:
\begin{eqnarray}
N_{pert}(\kappa) = K \cdot \left(\alpha_s(\kappa)\right)^b
 \cdot \exp{\left( \frac{c} {\sqrt {\alpha_s(\kappa)}}\right)}
\cdot\left[ 1 + O(\sqrt{\alpha_s}) \right ]
\label{eqn:mul_webber}
\end{eqnarray}
\begin{eqnarray}
b=\frac{1}{4}+\frac{2}{3}\frac{n_f}{\beta_0}
           \left(1-\frac{C_F}{C_A}\right)  ~~;~~
c=\frac{\sqrt{32C_A\pi}}{\beta_0}  ~~;~~
\beta_0=11-\frac{2}{3}n_f ~~.
\nonumber
\end{eqnarray}
A first and a second order $\alpha_s$ 
have been used with this expression with the number of active flavours,
$n_f$, equal to five.
An alternative prediction has been given in \cite{mult_khoze} using
the limited spectrum approach:
\begin{eqnarray}
N_{pert} = K \cdot \Gamma(B)\left (\frac{z}{2}\right)^{1-B} I_{1+B}(z)
\label{eqn:mul_khoze}
\end{eqnarray}
\begin{eqnarray}
B = \frac{33 + 2/9 n_f}{33 - 2 n_f}  ~~~;~~~
z= \log \frac{\kappa^2}{\Lambda^2} \gamma_0 ~~~;~~~ \gamma_0 = \sqrt{\frac{2}{\pi}
C_A \alpha_s(\kappa)} \nonumber
\end{eqnarray}
Here a first order $\alpha_s$ has always been used 
with $n_f$ taken as three \cite{khoze_privat}. 
$\Gamma$ is the Gamma-function and $I_B$ the modified Bessel-function.
$K$ is a non-perturbative scale factor.
The QCD scale parameter $\Lambda$ enters into the definition of
$\alpha_s(\kappa^2/\Lambda^2)$ \cite{colliderphysics}.
The numerical values of $K$ and $\Lambda$ are not expected to be the
same in Eqns.~\ref{eqn:mul_webber} and \ref{eqn:mul_khoze} as
different approximations are used.
Finally:
\begin{eqnarray}
r(\kappa)&=& \frac{C_A}{C_F} ( 1 - r_1 \gamma_0 -r_2 \gamma_0^2 )
\label{eqn:mueller_r}
\end{eqnarray}
with:
\begin{eqnarray*}
r_1 &=& \frac{1}{6}
\left (1 + \frac{n_f}{C_A} - \frac{2 n_f C_F}{C_A^2} \right)
~~;~~
r_2 = \frac{r_1}{6} 
          \left ( \frac{25}{8}
          -\frac{3}{4}\frac{n_f}{C_A} - \frac{n_f C_F}{C_A^2}
          \right ) 
\end{eqnarray*}
is the perturbative prediction \cite{mueller2}
for the multiplicity ratio in back-to-back
gluon to back-to-back quark jets.
The terms proportional to $r_1$ ($r_2$) correspond to the NLO (NNLO)
prediction.
Numerically they correspond to corrections of about 8\% and 1\%  respectively.
The smallness of the higher order corrections indicates that the perturbative
series of the gluon-to-quark multiplicity ratio converges rapidly.

The fits represent the data well.
The fit range has not been extended to too small scales as here
a contribution of initial two-jet events might bias the
multiplicities to lower values.
Parameters of the fits for this specific choice of scale and jet
selection are given in Tab. \ref{fit_mehrere}.
No estimate of systematic error is given as this analysis is intended
to be mainly qualitative.
The fit parameters should not be compared directly to
those parameters usually obtained from overall events in \epem\ annihilation.
The normalization factor, $K$, differs strongly due to the differences
in the multiplicity in jets and overall events. Furthermore, the
introduction of non-perturbative offsets leads to a strong reduction of
the values of the effective scale parameter $\Lambda$. 
This is also observed if the \epem\ multiplicity
is fitted including an offset term, which could be reasonable in
this case also.

\begin{table}[t,h,b]
\begin{center}
\newcommand\rup[1]{\raisebox{1.5ex}[-1.5ex]{#1}}
\newcommand\rdw[1]{\raisebox{-1.5ex}[-1.5ex]{#1}}
\newcommand\mc[2]{\multicolumn{#1}{#2}}
\begin{tabular}{|l|c|c|}
\hline
 \rdw{Parameter} &     $N_{pert}$ from      & $N_{pert}$ from      \\
                 &     Eqn.~\ref{eqn:mul_webber}& Eqn.~\ref{eqn:mul_khoze} \\
\hline
$\Lambda[$GeV$]$ &     0.032 $\pm$ 0.011    &  0.011 $\pm$ 0.004   \\
K                &     0.005 $\pm$ 0.001    &  0.12  $\pm$ 0.02    \\
$\cacf  $        &     2.12  $\pm$ 0.10     &  2.15  $\pm$ 0.10    \\
$N_0^q$          &     2.82  $\pm$ 0.14     &  3.12  $\pm$ 0.20    \\
$N_0^g$          &     0.73  $\pm$ 0.21     &  1.43  $\pm$ 0.31    \\
$\chi^2$/n.d.f.  &     0.61                 &  0.65                \\
\hline
\end{tabular}
\caption[]{Results of the fits of the quark and gluon jet multiplicities
as a function of $\kappa$.}
\label{fit_mehrere}
\end{center}
\end{table}

Using an identical scale definition for quark and gluon jets also
allows the gluon-to-quark jet multiplicity ratio to be directly evaluated as
function of this scale.
Fig. \mbox{\ref{qgcmp1}b)} shows this ratio 
as calculated from data and the fits
as function of the hardness scale as well as the ratio of the
slopes of the fits.
The ratio of the multiplicities increases from about 1.15 at small
scale to about 1.4 at the highest scales measured.
The measurement \cite{cleo} performed in $\Upsilon(1S) \rightarrow
\gamma {gg}$ decays at small 
scale\footnote{Half of the $gg$ invariant 
mass is taken as the equivalent scale.},
and of ``inclusive'' gluons \cite{opal} at large scale, agree quite well with
the expectation from the fits. 
The corresponding hardness scale for the data at the highest scale 
\cite{opal} has been estimated
from the average gluon energy and the angle cuts given in \cite{opal}.
The good agreement of the ``inclusive'' gluon measurement
also implies that angular ordering effects are relevant in this
case.

The ratio of the slopes for the different fits is almost 2 corresponding to
a colour factor ratio of $C_A/C_F = 2.12 \pm 0.10$, well compatible with
the QCD expectation.

The fits further indicate that for very small scale the multiplicity 
of quark jets is bigger than that of gluon jets.
Consequently the constant terms contributing to the
multiplicity due to the primary gluon or quark fragmentation
are larger for quarks (see Tab. \ref{fit_mehrere}).
The difference of these terms is about 2.
Taking the scale choice made in
\cite{aleph_qgmult} leads to about a 20\% increase of the 
measured colour factor
ratio and a corresponding increase in the difference of the non-perturbative
constants to 4.2. 

It is instructive here to estimate a lower limit for the difference of
the non-perturbative terms from the behaviour of the gluon and quark
fragmentation functions \cite{splitting_paper}. Due to leading particle
effects the fragmentation function of the quark outreaches the
fragmentation function of the gluon at high values of $x_E$. Taking the
shape of the gluon fragmentation function as unbiased by the leading
particle effect and assuming the overall multiplicity of gluon jets
roughly as twice as big as of quark jets, one gets an estimate for the
lower limit of additional multiplicity in quark jets by integrating 
the difference between the quark and the
halved gluon fragmentation function in the $x_E$-region where the
fragmentation function of the gluon is below that of the quark.
This yields $N_0^q-N_0^g \geq 0.61 \pm 0.02$ from Y and 
$N_0^q-N_0^g \geq 0.58 \pm 0.05$ from so-called Mercedes events
\cite{splitting_paper}. 
It should be noted here, that the leading particle effect still
influences the multiplicity at even lower scaled hadron energies.
The region of small hadron energy contributes most to the multiplicity.
Therefore the estimated limit presumably is much smaller than the
actual value of $N_0$.

At first sight a difference of the constant terms of the order of $\sim$2 units
in charged multiplicity looks unexpectedly large. 
However, these constants also include the effects of the jet clustering.
Furthermore,
stable hadron production to a large extent proceeds via 
resonance decays, so that
the observed difference may only correspond to a difference of
about one primary particle.
The larger constant term for quarks compared to gluons
explains the different behaviour of the ratio of
multiplicities and the slope ratio in 
\mbox{Fig.~\ref{qgcmp1}b)}.

The observed behaviour would be expected from
non-perturbative effects of the fragmentation in the leading quark or gluon.
In the cluster fragmentation model, an additional gluon to quark-antiquark
splitting is needed in the fragmentation of a gluon compared to that
of a quark.

\subsection{Precise Determination of \cacf\ from Multiplicities in
Three-Jet Events \label{sect:3jet_result}}
The analysis presented so far, as in most other comparisons of quark and gluon
jet multiplicities, has the disadvantage of relying on the association
of (maybe low energy) particles to jets.
Clearly this involves severe ambiguities and specifically does not
consider  coherent
soft gluon radiation from the initial $q\bar{q}g$ ensemble.
This can be avoided and a precise measurement can be obtained 
by studying the dependence of the total charged multiplicity
in three-jet events as function of the quark and gluon scales.
In fact there is a definite MLLA prediction  \cite{DKT} for this multiplicity
$N_{q\bar{q}g}$:
\begin{eqnarray}
N_{q\bar{q}g} = \left[2 N_q(Y^*_{q\bar{q}}) + N_g(Y^*_{g}) \right]
\cdot (1+{\cal O}(\frac{\alpha_s}{\pi}))
\label{eqn:3mul_0}
\end{eqnarray}
with the scale variables:
\begin{equation}
Y^*_{q\bar{q}} = \ln \sqrt{ \frac{p_q p_{\bar{q}}} {2\Lambda^2} } =
\ln \frac{E^*}{\Lambda}
~~~,~~~
Y^*_g = \ln \sqrt{
    \frac{ (p_qp_g) (p_{\bar{q}}p_g) }
         {2\Lambda^2(p_qp_{\bar{q}} ) } } =
\ln \frac{ p_1^{\perp} } {2\Lambda}
~~~,
\end{equation}
$N_q(Y^*_{q\bar{q}})$ and $N_g(Y^*_{g})$ describe the scale dependence of the
multiplicity for quark or gluon jets, respectively.
$\Lambda$ is a scale parameter and the $p_{q, \bar{q}, g}$ 
are the four-momenta of the quarks
and the gluon.
The three-jet multiplicity depends on the
quark energy, $E^*$, in the centre-of-mass system of
the quark-antiquark pair and on the transverse momentum scale 
of the gluon, $p_1^{\perp}$.
For comparison with data, this is expressed  in \cite{khoze_ochs} as a
dependence on the measured multiplicity in $e^+e^-$ events, 
$N_{e^+e^-}$, and 
the colour factor ratio
as given in Eqn.~\ref{eqn:mueller_r}.
In addition,  we again choose to add a constant term, $N_0$, to account
for differences in the fragmentation of quarks and gluons
as discussed above.
Thus, omitting correction terms:
\begin{eqnarray}
N_{q\bar{q}g} =                   N_{e^+e^-}(2E^*) +
                    r(p_1^{\perp}) \left \{ \frac{1}{2}
                    N_{e^+e^-}(p_1^{\perp}) - N_0 \right \}
~~~.
\label{eqn:3mul}
\end{eqnarray}
Although at  first sight this appears to be the incoherent sum of the
multiplicity of the two quark jets
and the gluon jet, this formula includes coherence effects in the
exact definition of the scales of the $N_{e^+e^-}$ terms \cite{khoze_privat}.
Nevertheless, subtracting the non-perturbative term $N_0$ within the 
curly brackets gives a physical interpretation for $N_0$ as the
additional multiplicity in quark jets due to the leading particle
effect, which is contained in the measured $N_{e^+e^-}$ and has to be
subtracted to get the gluon contribution to the multiplicity.

\begin{figure}[ptb]
\begin{center}

\begin{minipage}[t]{224pt}
 \epsfig{file=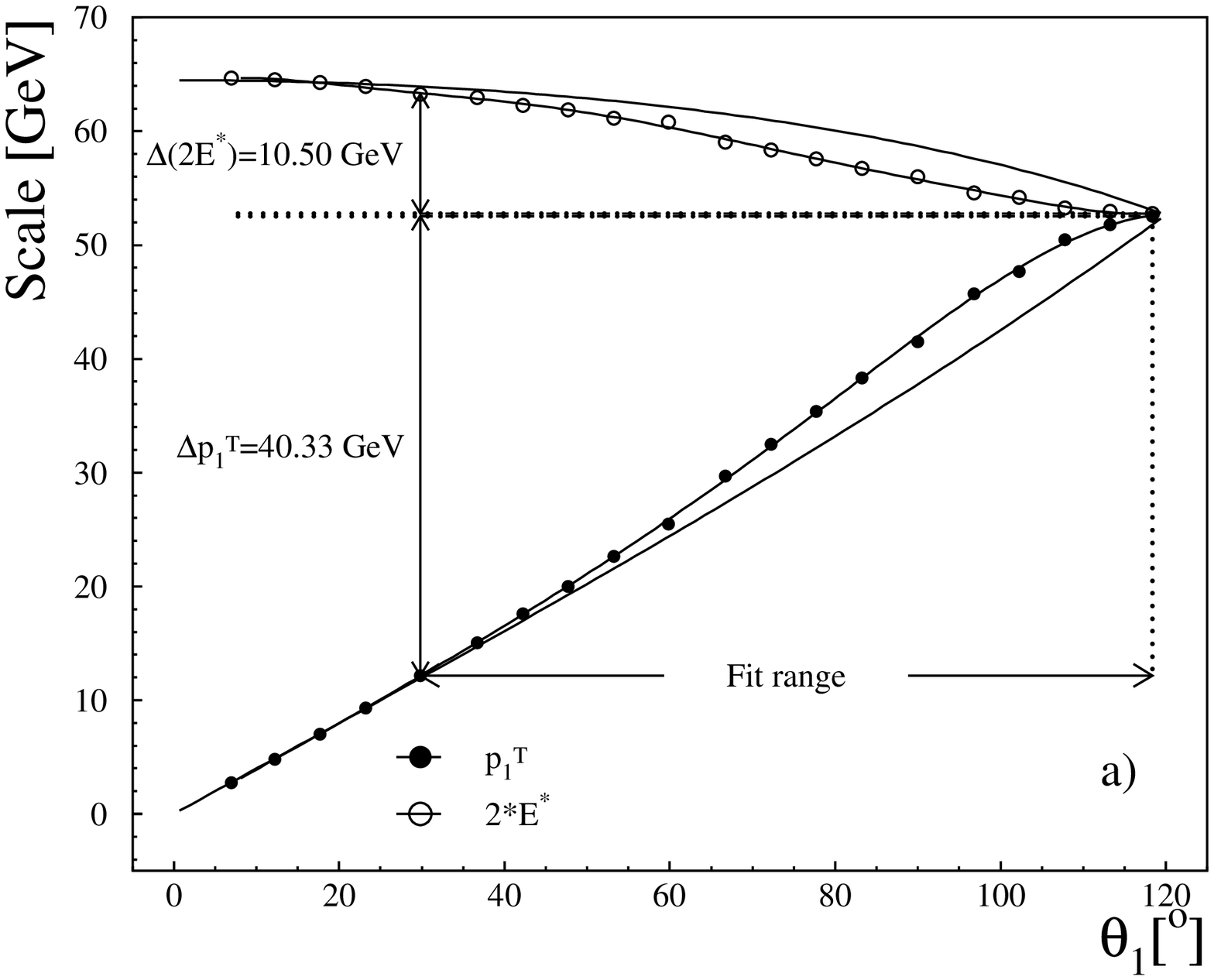,width=220pt}
\end{minipage}
\hfill
\begin{minipage}[t]{224pt}
\epsfig{file=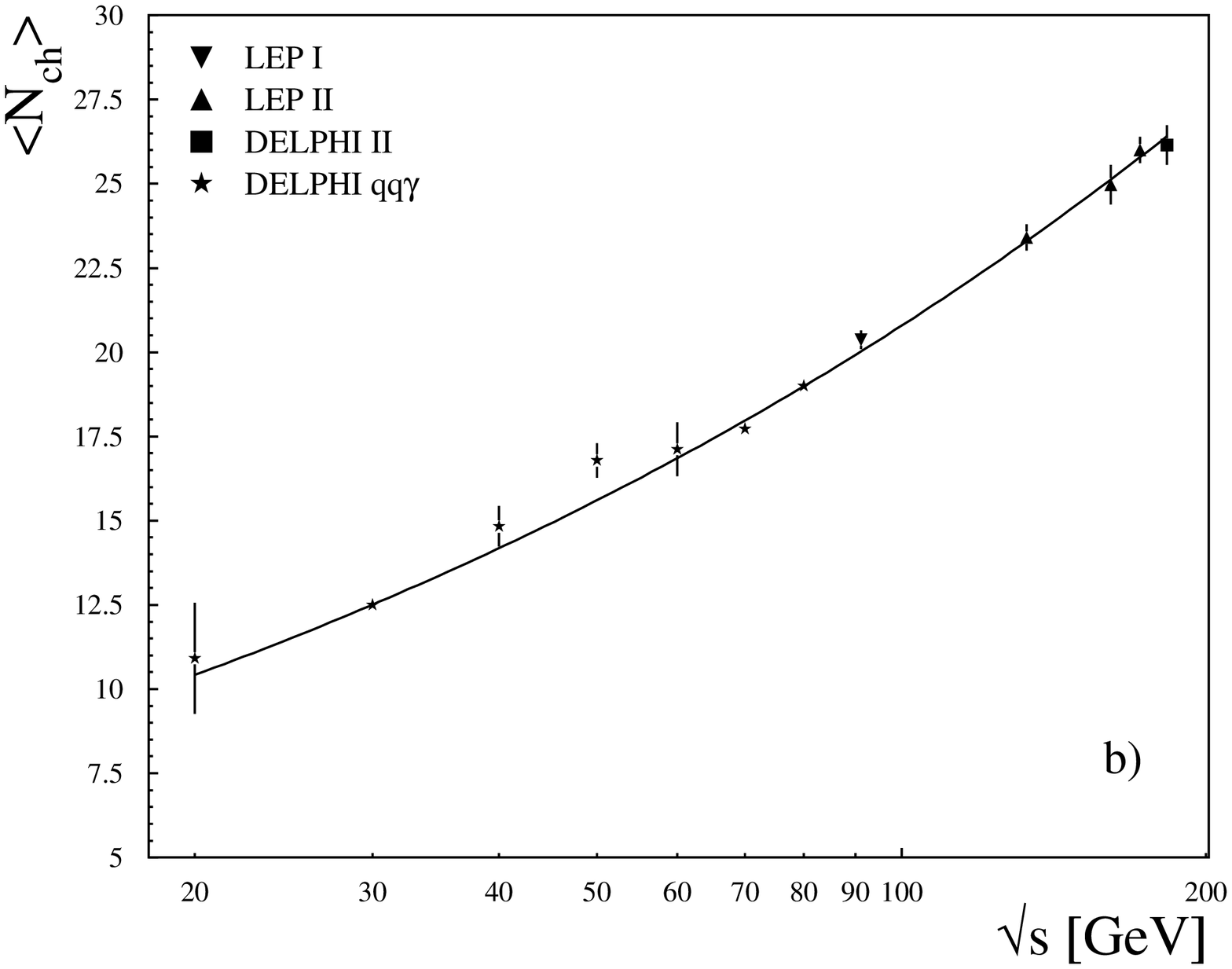,width=220pt}
\end{minipage}

\begin{minipage}[b]{224pt}
\epsfig{file=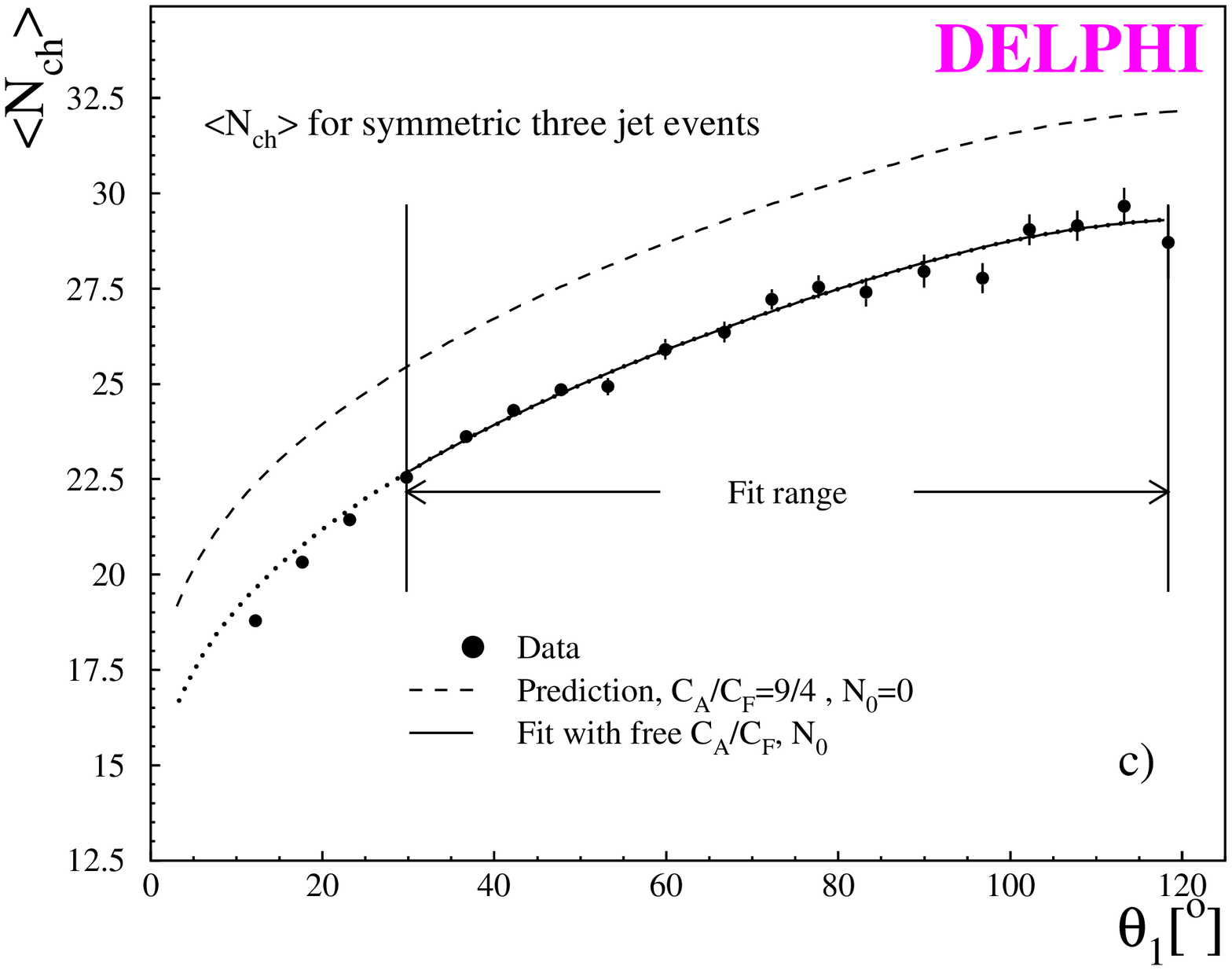,width=220pt}
\end{minipage}
\hfill
\begin{minipage}[b]{224pt}
\epsfig{file=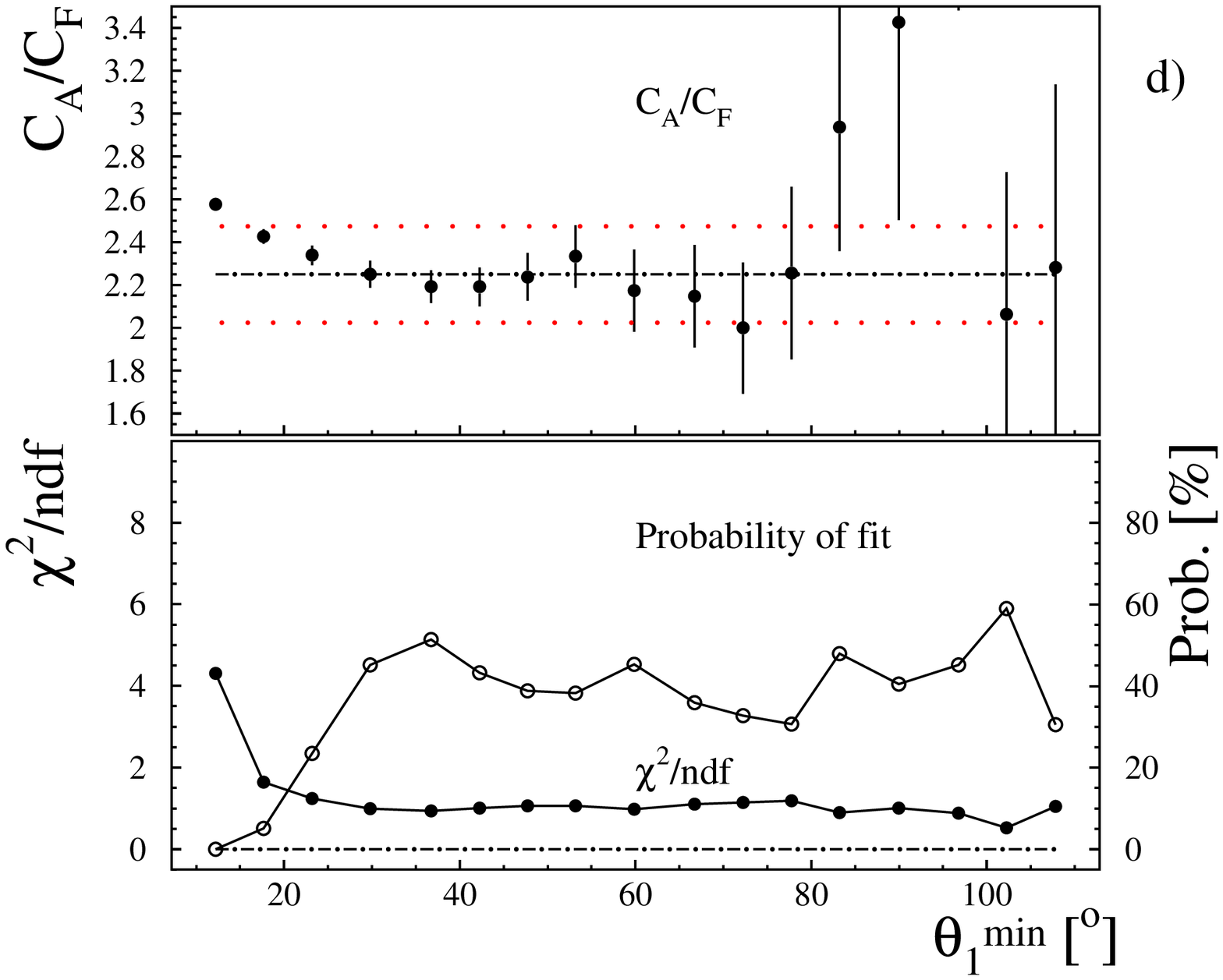,width=220pt}
\end{minipage}
\caption[]
{\label{result_plot}
a)Variation of the scales $2E^*$ and $p_1^{\perp}$ as function of the opening angle
$\theta_1$ in symmetric three-jet events. 
The functions are the analytic expectation.
The points include a correction (calculated with JETSET~7.3)
for the cases where the gluon forms the most
energetic jet. The lines matching the points are polynomials
fitted to obtain continuous values.  \\
b) Charged hadron multiplicity as a function of the centre-of-mass
energy of the $q \bar q $-pair fitted with the perturbative predictions Eqs. 
\ref{eqn:mul_webber} or \ref{eqn:mul_khoze}.\\
c) Charged hadron multiplicity in symmetric three-jet events as a
function of the opening angle. The dashed curve is the 
prediction using the ansatz
Eqn.~\ref{eqn:3mul} setting \cacf\ to its default value and omitting
the constant offset, $N_0$.
The full curve is a fit of the full ansatz \mbox{Eqn. \ref{eqn:3mul}} 
to the data treating
\cacf\ and $N_0$ as free parameters.\\
d) 
Stability of the result for \cacf\ against variation of the
smallest opening angle used in the fit as well as 
$\chi^2/N_{df}$ and the $\chi^2$ probability of these fits. The
dash-dotted horizontal line in the upper half shows the QCD expectation for 
\cacf\ with the dotted lines representing variations of $\pm 10\%$.\\
The DELPHI data of Fig. \ref{result_plot} b) and c) will be made available in
the Durham/RAL database \cite{durhamdb}.
}
\end{center}
\end{figure}

In principle Eqn.~\ref{eqn:3mul} still requires the determination of the
quark-antiquark and gluon scales independently.
However, in symmetric Y-type events (see Fig. \ref{eventtopol}b))
both scales can be expressed as functions of the opening angle $\theta_1$
only by initially assuming that the gluon jet is not the most energetic one.
\mbox{$E^{*2}\propto E_q E_{\bar{q}} \sin^2 \theta_3/2$} for this
type of event is almost constant (see upper full curve in Fig.
\ref{result_plot}a)) at fixed centre-of-mass energy.
However, $p_1^{\perp}$, increases approximately linearly with the opening
angle
as it is proportional to the gluon transverse momentum.
As the multiplicity change corresponding to the change of $E^*$
corresponds only to about $-2$,
the $\theta$ dependence of the three-jet multiplicity therefore mainly measures the
scale dependence of the multiplicity of the gluon jet.

In a fraction of the events (which strongly increases with opening angle)
the gluon jet is the most energetic jet.
This can be corrected for in different ways
when fitting Eqn.~\ref{eqn:3mul} to the data
using Monte Carlo simulation.
Assuming an approximately logarithmic increase of the multiplicity
with scale, which is 
well supported by the data, the average scale at a given opening angle
can be expressed as the geometric mean of the cases where the gluon
initiates the most energetic jet and where it does not.
These corrected scales are shown as the points in Fig. \ref{result_plot}a).
The correction first increases with the opening angle but then decreases again
and vanishes for fully symmetric events.
Alternatively, the fraction of events when the  gluon initiates the most
energetic jet can 
be considered separately in Eqn.~\ref{eqn:3mul}.

To obtain information on the colour factor ratio $C_A/C_F$, the scale dependence
of the three-jet multiplicity has to be compared to the multiplicity
in all $e^+e^-$ events.
This has been chosen to be taken from the DELPHI measurements with
hard photon radiation for energies below the Z mass and at 184 GeV \cite{delphi184}
and the LEP combined measurements at the intermediate energies \cite{lepcombi}.
For studies of systematic errors,
data from lower energy $e^+e^-$ experiments \cite{epemdata}
have also been used.
The DELPHI multiplicities in events with hard photon radiation
have been extracted as described in \cite{gluon_paper_1,salva}, but using
the full statistics now available.
Small energy dependent corrections ($2-4\%$) to the  $e^+e^-$ multiplicities
were applied to correct for the varying contribution of b quarks.
The  multiplicities  obtained were fitted with the perturbative predictions,
Eqs. \ref{eqn:mul_webber} or \ref{eqn:mul_khoze}, see Fig \ref{result_plot}b).
Both calculations describe the data equally well. 
The parameters of the fits are given in the upper part of Tab. \ref{finalfit}.

\begin{table}[t,h,b]
\begin{center}
\newcommand\rup[1]{\raisebox{1.5ex}[-1.5ex]{#1}}
\newcommand\rdw[1]{\raisebox{-1.5ex}[-1.5ex]{#1}}
\newcommand\mc[2]{\multicolumn{#1}{#2}}
\begin{tabular}{|l|c|c|c|}
\hline
 \rdw{Parameter} &     $N_{pert}$ from \cite{webber}
                                            & $N_{pert}$ from \cite{mult_khoze}
                                                                   &relevant\\
                 &     (Eqn.\ref{eqn:mul_webber})
                                            & (Eqn.\ref{eqn:mul_khoze})             &data\\
\hline
$\Lambda$        &     0.275 $\pm$ 0.070    &  0.061 $\pm$ 0.015   &data from\\
K                &     0.026 $\pm$ 0.003    &  0.606 $\pm$ 0.062   &$e^+e^-$ and $q\bar{q}\gamma$\\
$\chi^2$/n.d.f.  &     1.180                &  1.183               &      \\
\hline
\hline
$C_A/C_F$        &     2.251 $\pm$ 0.063    &  2.242 $\pm$ 0.062   &data from\\
$N_0$            &     1.40 $\pm$ 0.10      &  1.40 $\pm$ 0.10   &symmetric\\
$\chi^2$/n.d.f.  &     0.998                &  1.004             &3 jet events\\
\hline
\end{tabular}
\caption[]{Result of the fits of the \epem\ multiplicity (upper part)
and the three-jet event multiplicity (lower part).}
\label{finalfit}
\end{center}
\end{table}

The measured, fully corrected multiplicity in all symmetric three-jet events as
function of the opening angle is shown in Fig. \ref{result_plot}c).
A strong increase of the multiplicity from values of around 18 for
small opening 
angle to about 29 at opening angles of 120$^\circ$
(corresponding to fully symmetric events) is observed.
Omitting the non-perturbative term, $N_0$, in Eqn.~\ref{eqn:3mul} and setting
\cacf\ to its expected value predicts a similar increase
in multiplicity over this angular range (dashed curve in \ref{result_plot}c)).
The prediction is however higher by about three units of charged multiplicity.
This discrepancy is expected from the previously obtained result
due to differences in the fragmentation of the leading quark or gluon.

At small angles the difference between the primary QCD expectation and the
measurement increases.
Studies using Monte Carlo models have shown that this is mainly due to 
genuine two jet events which have been clustered as
symmetric three-jet events.
The models indicate that this contribution becomes small for angles above
30$^{\circ}$.

Fitting the full ansatz \ref{eqn:3mul} to the three-jet multiplicity
data at
angles $\theta \geq 30^{\circ}$, using the two parameterizations in Eqs.
\ref{eqn:mul_webber} and \ref{eqn:mul_khoze} of the
multiplicity in $e^+e^-$ events with their parameters fixed as given
in Tab. \ref{finalfit} but varying $C_A/C_F$ and $N_0$, yields:
\begin{eqnarray}
\label{eqn:res_webber}
\frac{C_A}{C_F} &=& 2.251 \pm 0.063\\
\label{eqn:res_khoze}
\frac{C_A}{C_F} &=& 2.242 \pm 0.062.
\end{eqnarray}
The result confirms with great precision the QCD expectation
\cite{brodskygunion} that the ratio of the radiated multiplicity
from gluon and quark jets is given by the colour factor ratio
\cacf. This result also implies that the proportionality of the
number of gluons to hadrons \cite{brodskygunion} e.g.~Local Hadron
Parton Duality (LPHD) \cite{lphd} applies extremely precisely if
only the radiated gluons from a quark or gluon are considered.

The offset term $N_0$  is bigger
if only b-depleted events are used.   The central result for
\cacf, however, remains unchanged within errors. This is due to
the fact that \cacf\ is measured from the change of multiplicity in
three-jet events with opening angle and not from the absolute
multiplicity.

The correctness of the ansatz Eqn.~\ref{eqn:3mul} and the
bias introduced by two-jet events at small $\theta_1$, 
were further checked by varying the lowest angle used in the fit.
The resulting value for $C_A/C_F$, the $\chi^2/N_{df}$ and the
$\chi^2$ probability of the fit are shown in \mbox{Fig. \ref{result_plot}d).}
It is observed that for  $\theta_1 > 30 ^{\circ}$ satisfactory fits are
obtained.
For this angular range the fitted value of $C_A/C_F$ is stable within errors.

Systematic uncertainties of the above  result for the colour factor
ratio due to uncertainties in the three-jet multiplicity data as well as
in the parameterization of the \epem\ charged multiplicity
and in the theoretical predictions are considered.
To obtain systematic errors interpretable like statistical errors, 
half the difference in the 
value obtained for \cacf\ when a parameter 
is modified from its central value 
(see below) is quoted 
as the systematic uncertainty.
All relative systematic errors are collected in Tab. \ref{systematic}.

\begin{table}[t,h,b]
\begin{center}
\begin{tabular}{|rl|c|c|c|c|}
\hline
 & Source & Sys. error    &     combined           &  combined     &  total \\
\hline
\hline
& \multicolumn{4}{l|}{Experimental uncertainties }&\multicolumn{1}{c|}{\ }  \\
\cline{1-5}
1. & Min. particle momentum          &   $\pm~ 0.42 $~\% & & & \\
2. & Min. angle of jet w.r.t. beam   &   $\pm~ 0.38 $~\% & & & \\
3. & Min. number of tracks per jet   &   $\pm~ 0.02 $~\% & $ \pm~0.58$\% & & \\
4. & Corr. for gluon in jet 1        &   $\pm~ 0.11 $~\% & & & \\
\cline{1-4}
5. & jet algorithms                  & & $\pm~ 1.39 $~\% &\raisebox{1.5ex}[-1.5ex]{$\pm~ 3.55$~\%} & \\
\cline{1-4}
6. & \epem\ data sets                 &  $\pm~ 0.90 $~\% & & & $\pm 5.52 \%$ \\
7. & Fit function                    &   $\pm~ 0.02 $~\% & $\pm~ 3.21 $~\% & & \\
8. & binning and range of fit        &   $\pm~ 3.08 $~\% & & & \\
\cline{1-5}
& \multicolumn{4}{l|}{Theoretical  uncertainties }&\multicolumn{1}{c|}{\ }  \\
\cline{1-5}
9. & Variation of $n_f$              &   $\pm~ 1.51 $~\% & & & \\
10.& Calculation in 1st/2nd order    &   $\pm~ 3.95 $~\% & &
                                                            $\pm~ 4.23$\% & \\
11.& Setting $C_A$ fixed             &   $\pm~ 0.08 $~\% & & & \\
\hline
\end{tabular}
\caption[]{Systematic uncertainties  on \cacf\ as derived from three-jet event
multiplicities }
\label{systematic}
\end{center}
\end{table}

Results for $\cacf$
obtained from the individual data sets corresponding to the different
years of data-taking 
as well as from b-depleted events
were found to be fully compatible within the statistical
error.
To estimate uncertainties in the three-jet multiplicity the following
cuts which are  sensitive to  misrepresentation of the data by the Monte Carlo
simulation have been varied.

\begin{enumerate}

\item {
Cut on the minimal particle momentum:
}\\
the cut on the minimal particle momentum has been lowered from 400 MeV
to \mbox{200 MeV} and raised to 600 MeV.
\item {
Minimum angle of each jet with respect to the beam axis:
}\\
this cut has been increased from 30$^\circ$ to 40$^\circ$ to test for a 
possible bias due to the limited angular acceptance.

\item {
Minimum number of particles per jet:
}\\
the minimum number of particles per jet has been increased from 2 to 4 in
order to reject events which may not have a clear three-jet structure.

\item {Correction for gluon in leading jet:
}\\
both methods of correction were compared
to account for gluons in the most energetic jet.
Furthermore the requirements for the mapping of the parton to the hadron level
for defining the gluon jet have been varied.

\end{enumerate}

To check the stability of the result for different choices of jet
algorithms the results obtained for a large sample of events 
generated with JETSET have been compared with:
\begin{enumerate}
\setcounter{enumi}{4}
\item {Alternative jet algorithms:
}\\
the angular ordered Durham algorithm, LUCLUS without particle reassignment,
JADE and Geneva \cite{jetalgos} were applied alternatively to Durham on
a large statistics Monte Carlo sample.
The results for Durham, angular ordered Durham and LUCLUS agree
reasonably.
The spread among the results was taken as error.
The JADE and Geneva algorithm which are known to tend to form
so-called junk jets \cite{jetalgos} show stronger deviations.
\end{enumerate}

The following systematic uncertainties arise from uncertainties in the
experimental input other than from the three-jet multiplicities
and from choices made for the fits of $N_{\epem}$.
These uncertainties are considered as experimental systematic uncertainties.

\begin{enumerate}
\setcounter{enumi}{5}
\item {
Input of parameterization of  $N_{e^+e^-}(\sqrt{s})$:
}\\
to estimate the influence of an uncertainty in  $N_{e^+e^-}$,
different choices of input data were compared:
\begin{itemize}
\item DELPHI multiplicities for 184 GeV and
      from Z decays with hard photons combined with
      LEP data for 90 GeV $<\sqrt{s}<$ 180 GeV ;
\item DELPHI multiplicities from Z decays with hard photons;
\item \epem\ data taken at low centre-of-mass energies
{(TASSO, TPC, MARK-II, HRS, AMY)};
\item all available \epem\ data between 10~GeV and 184~GeV
{(TASSO, TPC, MARK-II, HRS, AMY, LEP~combined, DELPHI)}.
\end{itemize}

\item {
Choice of prediction used for fit:
}\\
the fit functions \ref{eqn:mul_webber} and \ref{eqn:mul_khoze} were 
used alternatively.
For consistency here $n_f=5$ and a second order $\alpha_s$ was used.
\item {
Variation of the fitted range:
}\\
the lower limit of the angular range used in the fit was varied
between $24^\circ$ and $36^\circ$ as well as changing half the bin width,
$\epsilon$, from 2.5$^\circ$ to 5$^\circ$.

\end{enumerate}

Finally, systematic errors due to uncertainties in the theoretical prediction
were considered.

\begin{enumerate}
\setcounter{enumi}{8}
\item {
Variation of $n_f$:
}\\
the number of active quarks, $n_f$, \cite{colliderphysics}
relevant for the hadronic final state
is uncertain.
$n_f$ therefore has been varied from 3 to 5.

\item {
Order of calculation (LO - NNLO):
}\\
the prediction
$r(\kappa)$ (Eqn.~\ref{eqn:mueller_r}) has been calculated for
back-to-back quarks or gluons.
As the jets are well separated it is expected to  apply for this analysis also.
When the gluon recoils with respect to the quarks the prediction is exact.
In addition coherence effects (angular ordering) are taken into
account in the definition of the scales $E^*$ and $p_1^{\perp}$.

As the coupling for the triple-gluon vertex is bigger than the
coupling of
all other vertices it is clear that the correction will lower
the gluon-to-quark multiplicity ratio
as in the case of Eqn.~\ref{eqn:mueller_r}.
The validity of the correction \cite{mueller2} is therefore assumed for
the whole range of angles considered. Conservatively, half of the
difference obtained with the lowest order prediction $r=\cacf$ and the NNLO
prediction is considered as systematic uncertainty.
A leading order $\alpha_s$ was used for the lowest order prediction
and a second order $\alpha_s$ in the other case.
Considering that in the three-jet events mainly the gluon scale
$p_1^{\perp}$ is varied,
the resulting error estimate agrees with that given in
Eqn.~\ref{eqn:3mul_0}.

\item {Quantities influencing \cacf:}\\
for the central result, \cacf\ has been assumed variable in  Eqn.~\ref{eqn:mueller_r}
only.
The stability of the result was checked by also leaving $C_A$ variable
in some or all of the parameterizations of $\alpha_s$ and $N_{\epem}$.
\end{enumerate}

To check in how far the offset term $N_0$ is constant, $N_0$ has been
extracted for each $\theta_1$-bin individually fixing $C_A$ to its
default value. The individual results are consistent with the average
value and no trend is observed.

Alternatively to Eqn.~\ref{eqn:3mul}, Eqn.~\ref{eqn:3mul_0} has been
fitted to the data, where the \mbox{${\cal O}(\alpha_s)$} correction factor has
been parameterized as \mbox{$(1+c\alpha_s(p_1^\perp))$}. This leads to the same 
fit results for $C_A/C_F$ and $\chi^2$ as Eqn.~\ref{eqn:3mul}, which
implies that both corrections \mbox{$r(p_1^\perp)\cdot N_0$} and 
\mbox{$\left[2 N_q(Y^*_{q\bar{q}}) + N_g(Y^*_{g})
\right]\cdot c\alpha_s(p_1^\perp)$} 
as well as the values obtained for $\cacf$
agree within $\pm 1\%$.
It should, however, be stressed that the behaviour of the
fragmentation function requires the presence of a non-perturbative
offset term.

The prediction of the multiplicity ratio given by
\cite{dremin_nechaitilo} has been tried as an alternative to 
Eqn.~\ref{eqn:mueller_r}. 
Although this calculation takes recoil effects into account,
a non-perturbative offset term is still required.
The prediction differs by about 10\% from \cite{mueller2} in the NNLO term.
As it does not reproduce the colour factor ratio contained in the
fragmentation models which describe the data well,
it has not been applied in this analysis.

Averaging the results given in Eqns. \ref{eqn:res_webber} and
\ref{eqn:res_khoze} and adding in quadrature the systematic errors
summarized in Tab. \ref{systematic} gives the following final result:
\begin{eqnarray}
\frac{C_A}{C_F} = 2.246 \pm 0.062~(stat.) \pm 0.080~(syst.) \pm 0.095~(theo.) 
\end{eqnarray}
This result confirms the QCD expectation that gluon bremsstrahlung
is stronger from gluons than from quarks by the colour factor ratio
\cacf\ and is direct evidence for the triple-gluon coupling.

This measurement yields
the most precise result obtained so far for the colour factor ratio 
\cacf. Even the best
measurements from four-jet angular distributions \cite{4jet} suffer from
the relatively small number of four-jet events available.
Furthermore, many of these measurements specify
no theoretical systematic error as they so far rely on leading order
calculations.
It is remarkable that this measurement of \cacf\ is performed from
truly hadronic quantities, the charged multiplicities. 
Jets, i.e. partonic quantities only enter indirectly via the definition
of the scales $E^*$ and $p_1^{\perp}$.

\begin{figure}[tb]
\begin{center}
\epsfig{file=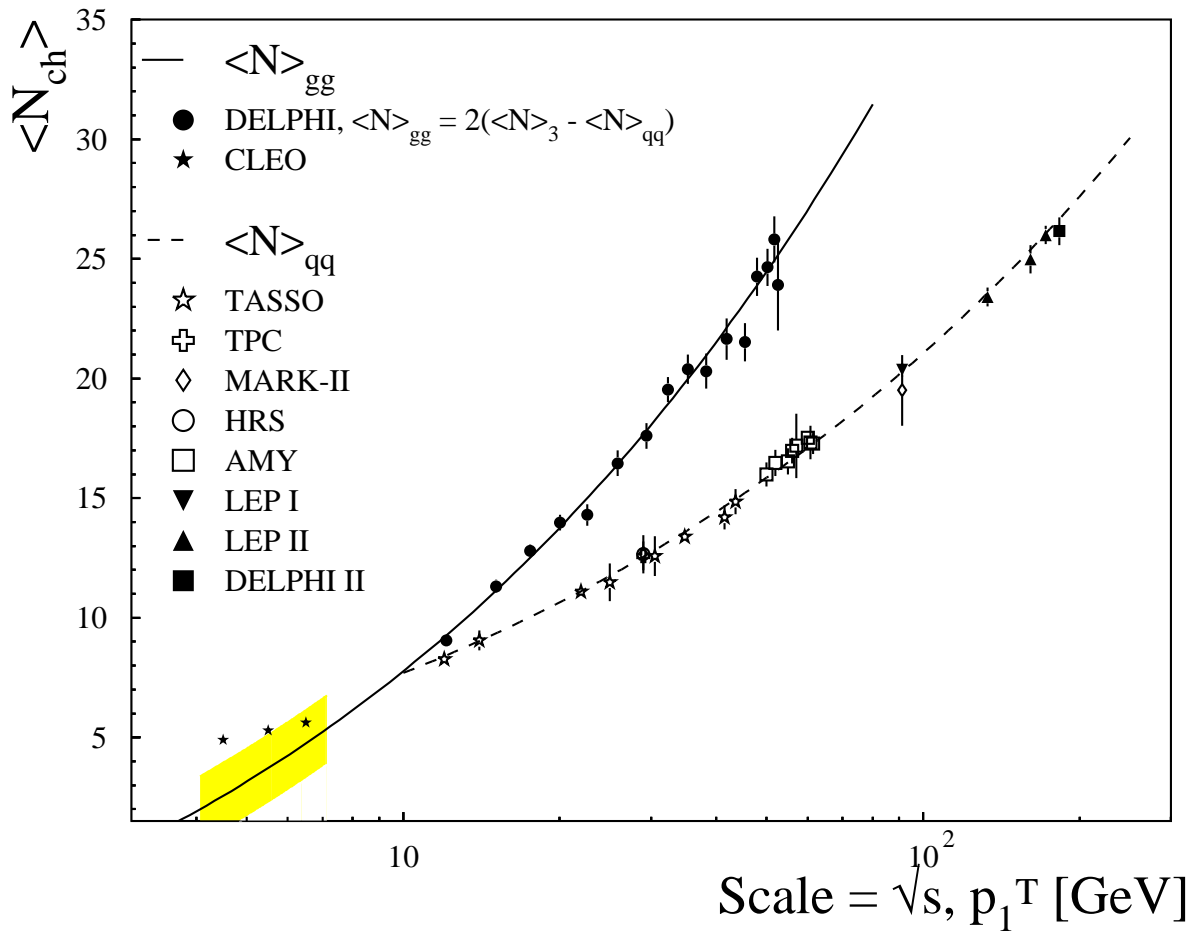,width=15.8cm}
\caption[summ_multplot]
{
\label{qq_gg_mult}
Comparison of the charged hadron multiplicity for an initial 
\qqbar\ and a $gg$ pair as
function of the scale.
The dashed curve is a fit according to Eqns.~\ref{eqn:mul_webber} or
\ref{eqn:mul_khoze}, the full line is twice the second term of
Eqn.~\ref{eqn:3mul}.
The grey band indicates the uncertainty due to the error of $N_0$.
The DELPHI $gg$ data will be made available in
the Durham/RAL database \cite{durhamdb}.
}
\end{center}
\end{figure}

In order to illustrate comprehensively the contents of the measurement
of the three-jet multiplicity we compare in Fig. \ref{qq_gg_mult} the
multiplicity corresponding to a $gg$ and a \qqbar\ final state.
The \qqbar\ multiplicity is taken to be the multiplicity measured in
\epem\ annihilation corrected for the ${\rm b\bar{b}}$ contribution as
described above.
The $gg$ multiplicity at low scale values is taken from the CLEO measurement
\cite{cleo}, for which no systematic error was specified.
At higher scale, twice the difference of the three-jet multiplicity and the
\qqbar\ term (the first term in Eqn.~\ref{eqn:3mul}) is interpreted as
the $gg$ multiplicity.
The $gg$ data should be extendable to higher energies by measuring the
multiplicity in $\rm p\bar{p}$ scattering as a function of the
transverse energy.
The dashed curve through the \qqbar\ points is a fit of the prediction
according to Eqns.~\ref{eqn:mul_webber} or \ref{eqn:mul_khoze}.
The $gg$ line is the perturbative expectation for back-to-back gluons
according to the second term of Eqn.~\ref{eqn:3mul}.
$N_0$ is taken from Eqn.~\ref{eqn:N0res}.
In principle $N_0$ is a property of the complete three-jet event,
so it is unclear if the subtraction of the full amount of
$N_0$ is justified in order to obtain the gluon jet multiplicity.
However, this only introduces a constant shift in the ``$gg$ event''
multiplicity,
the scale dependence of the gluon jet multiplicity remains unaltered.
The plot shows again that the increase of the 
$gg$ multiplicity with scale is about twice as big as in the \qqbar\ case,
illustrating the large gluon-to-quark colour factor ratio \cacf.

It is of interest to present also a dedicated measurement of the
non-perturbative parameter $N_0$. In order to obtain this value,
b-depleted events have been used.
A fit of the three-jet event multiplicity has then been performed with
$N_0$ as the only free parameter.
\cacf\ has been set to its default value.
The parameterization of the \epem\ multiplicity according to 
Eqn.~\ref{eqn:mul_webber}
uses the low energy \epem\ data as input.
The fit yields:
\begin{equation}
N_0 = 1.91 \pm 0.03 {\rm (stat.)} \pm 0.33 {\rm (syst.)}
\label{eqn:N0res}
\end{equation}
The systematic error was estimated as for \cacf. 
Furthermore a normalization error due to the
multiplicity in \epem\ events has been added in quadrature. This
error has been assumed to be given by the error of the precise average
multiplicity at the Z resonance \cite{multtab}. 
The actual
value of $N_0 \approx 2$  corresponds to about one primary
particle (see also section \ref{sect:cmp_result}). 
This is indeed a reasonable value which had already been
expected in \cite{brodskygunion}.

\section{Summary}
In summary, the dependence of the charged particle
multiplicity in quark and gluon jets on the transverse momentum-like 
scale has been investigated
and the charged hadron multiplicity in symmetric three-jet 
events has been measured as a function of the opening angle $\theta_1$.

The ratio of the variations
of gluon and quark jet multiplicities with scale agrees with 
the QCD expectation
and directly
reflects the higher colour charge of gluons compared to quarks.
This can also be interpreted as direct evidence for the triple-gluon
coupling, one of the basic ingredients of QCD.
It is of special importance that this evidence is due to very soft
radiated gluons and therefore complementary to the measurement of
the triple-gluon
coupling in four-jet events at large momentum transfer.

The increase of the gluon to quark jet multiplicity ratio with increasing scale
is understood as being due to a difference in the fragmentation of the leading
quark or gluon.
The simultaneous description of the quark and gluon jet multiplicities
with scale also 
supports the Local Parton Hadron Duality
hypothesis \cite{lphd} although
large non-perturbative terms for the leading quark or gluon are
responsible for the observed relatively small gluon to quark jet multiplicity
ratio.

Using the novel method of measuring the evolution 
of the multiplicity in symmetric three-jet events with their opening angle,
a precise  result for the colour factor ratio is obtained:
$$
\frac{C_A}{C_F} = 2.246 \pm 0.062~(stat.) \pm 0.080~(syst.) \pm 0.095~(theo.) 
$$
It is superior in precision to the best
measurements from four-jet events \cite{4jet}.
Finally it is remarkable that this measurement is directly performed from
truly hadronic quantities. Jets only enter indirectly via the definition
of the energy scale of the quark-antiquark pair and the transverse
momentum scale of the gluon.
These scales are calculated directly from the jet angles.

\newpage

\subsection*{Acknowledgements}
\vskip 3 mm
We would like to thank V.A. Khoze for his interest in this analysis and
many enthusiastic discussions and explanations.
We thank S. Lupia and W. Ochs for providing us with their program for
Eqn.~\ref{eqn:mul_khoze}.

 We are greatly indebted to our technical 
collaborators, to the members of the CERN-SL Division for the excellent 
performance of the LEP collider and to the funding agencies for their
support in building and operating the DELPHI detector.\\
We acknowledge in particular the support of: \\
Austrian Federal Ministry of Science and Traffics, GZ 616.364/2-III/2a/98, \\
FNRS--FWO, Belgium,  \\
FINEP, CNPq, CAPES, FUJB and FAPERJ, Brazil, \\
Czech Ministry of Industry and Trade, GA CR 202/96/0450 and GA AVCR A1010521,\\
Danish Natural Research Council, \\
Commission of the European Communities (DG XII), \\
Direction des Sciences de la Mati$\grave{\mbox{\rm e}}$re, CEA, France, \\
Bundesministerium f$\ddot{\mbox{\rm u}}$r Bildung, Wissenschaft, Forschung 
und Technologie, Germany,\\
General Secretariat for Research and Technology, Greece, \\
National Science Foundation (NWO) and Foundation for Research on Matter (FOM),
The Netherlands, \\
Norwegian Research Council,  \\
State Committee for Scientific Research, Poland, 2P03B06015, 2P03B03311 and
SPUB/P03/178/98, \\
JNICT--Junta Nacional de Investiga\c{c}\~{a}o Cient$\acute{\mbox{\rm i}}$fica 
e Tecnol$\acute{\mbox{\rm o}}$gica, Portugal, \\
Vedecka grantova agentura MS SR, Slovakia, Nr. 95/5195/134, \\
Ministry of Science and Technology of the Republic of Slovenia, \\
CICYT, Spain, AEN96--1661 and AEN96-1681,  \\
The Swedish Natural Science Research Council,      \\
Particle Physics and Astronomy Research Council, UK, \\
Department of Energy, USA, DE--FG02--94ER40817. \\

\end{document}